# WAVELET ANALYSIS OF SOLAR ACTIVITY

*Stefano Sello*
*Thermo – Fluid Dynamics Research Center*
*Enel Research*
*Via Andrea Pisano, 120*
*56122 PISA - ITALY*

*ABSTRACT*

Using wavelet analysis approach, the temporal variations of solar activity on time scales ranging from days to decades are examined from the daily time series of sunspot numbers. A hierarchy of changing complex periods is careful detected and related cycles compared with results from recent similar analyses. A general determination of the main Schwabe cycle length variations is also suggested on the basis of the wavelet amplitude distribution extracted from the local wavelet power map.

INTRODUCTION

The multiscale evolution of the solar magnetic fields is extremely complicated and this is also transferred to the Sun global activity. The spatial and temporal patterns appearing on the surface of the Sun reflect the underlying dynamo mechanisms which drive the turbulent plasma motions inside the convection region. There exist now many strong indications that the related dynamics is well described, at least for a limited range of scales, by a low dimensional chaotic system [1],[2],[3].
The traditional way to record the variations of solar activity is to observe the sunspot numbers which provide an index of the activity of the whole visible disk of the Sun. These numbers are determined each day and are calculated counting both isolated clusters of sunspots, or sunspot groups, and distinct spots. The current determination of the international sunspot number, Ri, from the Sunspot Index Data Center of Bruxelles, results from a statistical elaboration of the data deriving from an international network of more than twenty five observing stations. Despite sunspot numbers are an indirect measure of the actual physics of the Sun photosphere, they have the great advantage of covering various centuries of time. This property results very crucial when we focus the attention on complex variations of scales and cycles related to the solar magnetic activity.

Many previous works based on standard correlation and Fourier spectral analyses of the sunspot numbers time series, reveal a high energy content corresponding to the Schawbe (~11 years) and a lower energy long term modulation corresponding to the Gleissberg (~100 years) cycles. A clear peak is also evident near the 27 days solar rotation as viewed from Earth. More careful statistical analyses show that these periodicities are in fact strongly changing both in amplitude and in scale, showing clear features of transient phenomena and intermittency [1]. The non stationary feature of those time series is well pointed out by the temporal behaviour of the related high level variance ($2^{nd}$ order moment). Wavelet analysis offers an alternative to Fourier based time series analysis, especially when spectral features are time dependent. By decomposing a time series into time-frequency space we are able to determine both the dominant modes and how they vary with time. Multiscale analysis based on the wavelet approach, has now been successfully used for many different physical applications, including geophysical processes, bio-medicine, and flow turbulence [4]. A high development for astrophysical applications resulted as a consequence of an extension of the wavelet approach to unevenly sampled time series, proposed by Foster in 1996 [5].

There are relatively few and recent works documented in literature, related to applications of wavelet analysis to solar activity, the first one dated back to 1993 by Ochadlick [6]. In the paper the author used a wavelet approach to detect "subtle" variations in the solar cycle period from yearly means of sunspot numbers. Multiscaling and intermittency features have been well analysed through the use of a continuous Morlet wavelet transform on the monthly sunspot numbers time series in the work of Lawrence et al. (1995) [1]. More recently, different authors used longer daily sunspot numbers time series to investigate more accurately new subtle periodicities and their evolutions [7],[8].

The main aim of the present paper is to add a further contribution to the wavelet analyses of solar activity through the application of the Foster projection method (with the estimation of confidence levels), to unevenly sampled time series derived from daily international sunspot numbers (1818-1999) (SIDC) [9], and from less noisy daily group sunspot numbers (1610-1995) (Hoyt and Schatten) [10].

WAVELET ANALYSIS

Fourier analysis is an adequate tool for detecting and quantifying constant periodic fluctuations in time series. For intermittent and transient multiscale phenomena, the wavelet transform is able to detect time evolutions of the frequency distribution. The continuous wavelet transform represents an optimal localized decomposition of time series, $x(t)$, as a function of both time $t$ and frequency (scale) $a,$ from a convolution integral:

$$W(a,\tau) = \frac{1}{a^{1/2}} \int_{-\infty}^{+\infty} dt\, x(t)\psi^*(\frac{t-\tau}{a})$$

where $\psi$ is called an analysing wavelet if it verifies the following admissibility condition:

$$c_\psi = \int_0^{+\infty} d\omega\, \omega^{-1} \left|\hat{\psi}(\omega)\right|^2 < \infty$$

where:

$$\hat{\psi}(\omega) = \int_{-\infty}^{+\infty} dt\, \psi(t) e^{-i\omega t}$$

is the related Fourier transform. In the definition, $a$ and $\tau$ denote the dilation (scale factor) and translation (time shift parameter), respectively.
We define the local wavelet spectrum:

$$P_\omega(k,t) = \frac{1}{2 c_\psi k_0} \left|W(\frac{k_0}{k},t)\right|^2 \quad k \geq 0$$

where $k_0$ denotes the peak frequency of the analysing wavelet $\psi$.
From the local wavelet spectrum we can derive a mean or global wavelet spectrum, $P\omega(k)$:

$$P_\omega(k) = \int_{-\infty}^{+\infty} dt\, P_\omega(k,t)$$

which is related to the total energy E of the signal x(t) by:

$$E = \int_0^{+\infty} dk\, P_\omega(k)$$

The relationship between the ordinary Fourier spectrum $P_F(\omega)$ and the mean wavelet spectrum $P_\omega(k)$ is given by:

$$P_\omega(k) = \frac{1}{c_\psi k} \int_0^{+\infty} d\omega \, P_F(\omega) \left| \hat{\psi}\left(\frac{k_0 \omega}{k}\right) \right|^2$$

indicating that the mean wavelet spectrum is the average of the Fourier spectrum weighted by the square of the Fourier transform of the analysing wavelet ψ shifted at frequency k.

Here we used the family of complex analysing wavelets consisting of a plane wave modulated by a Gaussian (called Morlet wavelet) [11]:

$$\psi(\eta) = \pi^{-1/4} e^{i\omega_0 \eta} e^{-\eta^2/2}$$

where $\omega_0$ is the non dimensional frequency here taken to be equal to 6 in order to satisfy the admissibility condition. For a more comprehensive and detailed description of the wavelet formalism see references [11], [12].
Following Foster (1996) [5] here we consider an extension of the above wavelet formalism in order to correct handle irregularly sampled time series. The wavelet transform is viewed as a suitable weighted projection onto three trial functions giving the Weighted Wavelet Z transform and the Weighted Wavelet Amplitudes. For all the mathematical details of this formulation and its applications we refer to Foster (1996) and Haubold (1998) papers [5],[12].

Many applications of the wavelet analysis suffered from an apparent lack of quantitative evaluations especially by the use of arbitrary normalization and the lack of statistical significance test in order to estimate the reliability of results. Here we used power spectrum normalization and significance levels following the approach suggested by Torrence et al. (1997). We first assume an appropriate background spectrum and we suppose that different realizations of the considered physical process will be randomly distributed about this expected background spectrum. The actual spectrum is compared against this random distribution. In the present work we assumed as background spectrum a red noise spectrum modeled through a univariate lag-1 autoregressive process:

$$x_n = \alpha x_{n-1} + z_n$$

where $z_n$ is derived from a Gaussian white noise and α is the lag-1 autocorrelation here estimated by:

$$\alpha = \frac{\alpha_1 + \sqrt{\alpha_2}}{2}$$

where $\alpha_1$ and $\alpha_2$ are the lag-1 and lag-2 autocorrelations of the considered time series. The discrete normalized Fourier power spectrum of this red noise is:

$$P_k = \frac{1-\alpha^2}{1+\alpha^2 - 2\alpha\cos(2\pi k/N)}$$

and the following null hypothesis is defined for the wavelet power spectrum: we assume the red noise spectrum as the mean power spectrum of the time series; if a peak in the wavelet power spectrum is significantly (here at 95% confidence level) above this background spectrum, then it can be considered to be a real feature of the time series (see [11]).

SOLAR ACTIVITY: SUNSPOT NUMBERS

We considered here two records of solar activity: 1) the daily international number of sunspots from SIDC archive covering the time interval: 1818-1999 and consisting of 66474 observations; 2) the daily number of sunspots groups visible on the Sun surface between 1610 and 1995 recently made available by Hoyt and Schatten [10], and consisting of 111358 observations. These time series are irregularly sampled especially in the first part of the records. The importance of the last time series results in its more internally self-consistency and in its lesser noise level than that of the Wolf sunspot numbers. It uses only the number of sunspot groups observed, rather than groups and individual sunspots. The group sunspot numbers use 65941 observations from 117 observers active before 1874 not used by Wolf in its original time series. The noise contained in this time series is considerably less than the noise contained in the Wolf sunspot numbers. Of course the numerical computation of the wavelet spectrum is here very expensive and time consuming due to the high number of data contained in the record.
Moreover, recently Ballester et al. (1999) [8], analyzing the daily group sunspot numbers with a wavelet approach, detect local episodes of the periodicity near 158 days (varying from 140 to 170 days) around the maximum of solar cycle 2 and around the maxima of solar cycles 16-21. They claimed that the presence of that periodicity in the group sunspot numbers confirms that it is caused by a periodic emergence of magnetic flux. A near 154 days periodicity in the Sun was first reported on gamma ray flares, Rieger et al. 1984 [13], and in other related parameters but it does not seemed to be a persistent periodicity. It is well known that solar flares are huge explosions on the surface of the Sun, involving time scales of a few minutes and they are thought to be formed as magnetic fields structures related to sunspots are twisted, releasing energy by magnetic reconnections. During solar cycle 21 a periodicity between 152, 158 days in the occurrence rate of energetic flares was detected by the Solar Maximum Mission satellite. If we are able to confirm a temporal and frequency agreement in the wavelet analysis of the group sunspot numbers, we can support the emergence of magnetic flux in the solar photosphere as the common factor relating the two phenomena.

- Daily International Sunspot Numbers

Figure 1 shows the results of the wavelet analysis applied to the series of daily international sunspot numbers. The upper part shows the original time series in its natural units (red line) and the monthly smoothed sunspot numbers (green line) generally used for solar cycle predictions. Time is here expressed in years. The central part shows the amplitudes (WWA) of the wavelet local power spectrum in terms of an arbitrary colour contour map. Red higher values are strong energetic contributions to power spectrum, while blue lower values are weak energetic contributions. Horizontal time axis corresponds to the axis of time series and vertical scale (frequency) axis is for convenience expressed in log values of cycles per year$^{-1}$. Thus the range analyzed is between 148 years (value -5) and 2.5 days (value 5). The right part shows the mean global wavelet power spectrum (red line) obtained with a time integration of the local wavelet power spectrum, and the 5% significance level using a red noise autoregressive lag-1 background spectrum with $\alpha=0.96$ (green line). It is possible to distinguish two main regions corresponding to sharp Schwabe (~11 years) and broad Gleissberg (~100 years) with dominant energetic contribution above the red noise background spectrum. In particular it is well evident the complex evolution of the Schwabe cycle both for scales and amplitudes. The irregular finite extension in the map represents a continuous interaction of frequencies. It is interesting to note the presence of a non persistent significant region around a period of 28 years starting from 1950. In this map there are not clear evidences of episodes related to 158 days periodicity (value 0.8). In fact, for periods below 3 years there are not significant contributions in the global energy spectrum, suggesting dominant stochastic motions [1] with little exceptions. Note the complex shape regions around maxima of cycles related to 27 days solar rotation, as viewed from Earth.

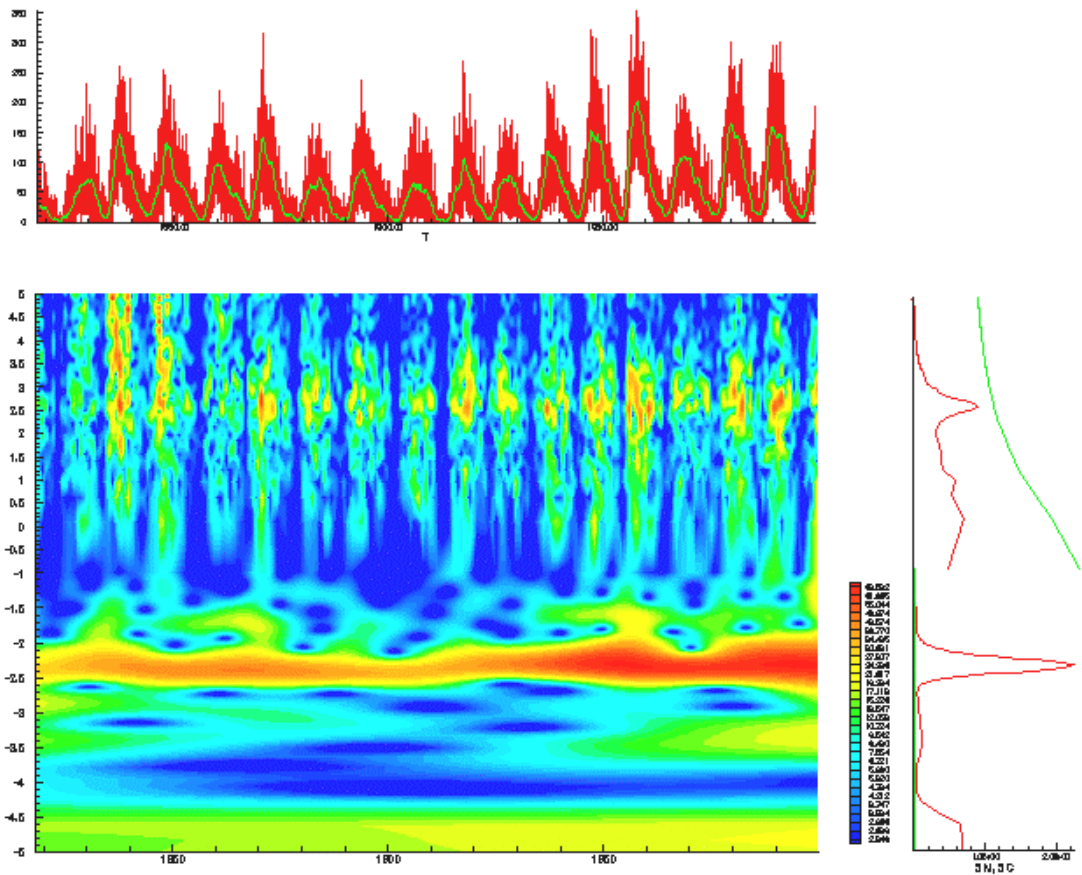

Figure 1

- Daily Group Sunspot Numbers

The original record of daily group sunspot numbers was limited between 1653 and 1995 because of the lack of reliability for observations before 1653. The total number of observation is then reduced to 104108. The lower global level of noise in this series is evidenced by a reduction of the variance of about 25%. On the other hand, here the red noise autoregressive lag-1 background spectrum was obtained with $\alpha=0.85$, i.e. with more features typical of white noise. Figure 2 shows the results of the wavelet computation in the same way of previous analysis.

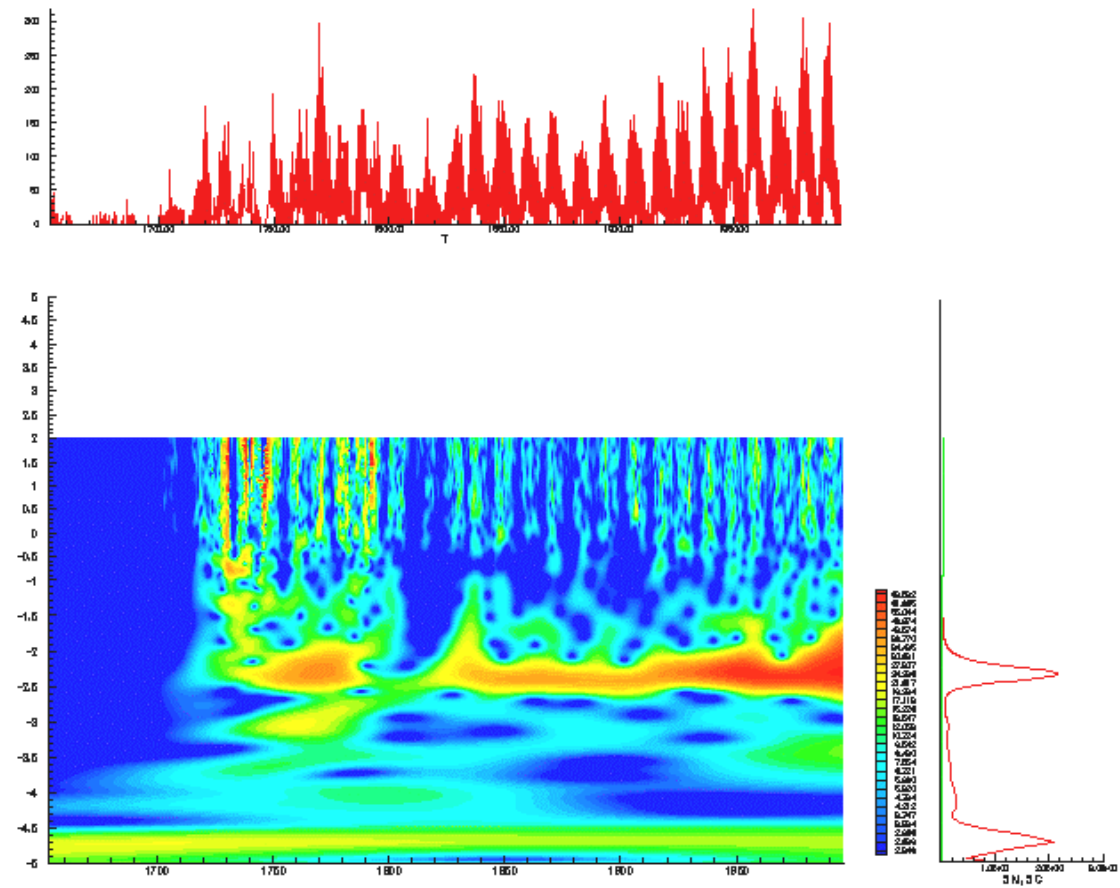

Figure 2

The high frequency range explored was limited up to 49 days period, due to computational restrictions. As in the previous analysis, only two dominant peaks were detected, corresponding to the Schwabe and the Gleissberg cycles. Here the longest cycle was detected more precisely, near 110 years. The interesting part of this wavelet analysis is related to the long Maunder minimum ending at 1715 and the next weaker minimum between 1795 and 1820. The evolution of the main cycle near 11 years, as derived from both the wavelet analyses, results quite similar. There is a strong increase of the energy contribution for recent cycles, with a continuous irregular variation of the region related to this periodicity, currently shifted toward high value frequencies. Moreover, a detailed inspection of the wavelet map between 140 and 170 days periods, shows again the absence of significant contributions localised near a 158 days periodicity for the group sunspot numbers. Thus, our wavelet analysis does not support the correlation between the periodicity detected in some high energy solar flares and the periodic emergence of magnetic flux associated to sunspot groups, as indicated in [8].

- Solar Cycle Length

The long term variations of the solar cycle length have been widely studied and its importance is mainly due to their suggested correlation to global climate [14].

Recently Mursula et al. [15], proposed a new method to determine the solar cycle length based on a difference between the median activity times of two successive sunspot cycles. The advantage of this method, with respect to a conventional approach, is that the median times are almost independent of how the minima or maxima are determined and thus the computation results more accurate. More recently, Fligge et al. [16], proposed a more objective and general cycle length determination using a continuous wavelet transform.

Here we propose an alternative general method to determine time variation of the main solar cycle, based on the amplitudes of the wavelet map for sunspot group numbers (Figure 2). The main idea is to extract a characteristic frequency (or period) for each time, from an irregular shaped region related to the main solar cycle. From the global wavelet power spectrum we selected a suitable range of frequencies, including the irregular region related to the Schwabe cycle: $\omega_0 = -3$ (20 years), $\omega_1 = -1.6$ (5 years). We then define a characteristic frequency of the wavelet map, at a given time t, as:

$$\tilde{\omega}(t) = \frac{\int_{\omega_0}^{\omega_1} d\omega\, \omega\, w(W(\omega,t))}{\int_{\omega_0}^{\omega_1} d\omega\, w(W(\omega,t))}$$

where $w(W(\omega,t))$ are proper weights derived from the intensity of the local wavelet power map. This characteristic frequency gives a wavelet based evaluation of the evolution of the periods related to the selected range of frequencies.

When we apply the above relation to the main Schwabe cycle we obtain the result shown in Figure 3. The central part of the Figure shows the behaviour of the characteristic periods or lengths of the solar cycle. For a comparison, we show also the solar cycle lengths obtained with the median method (blue line) (from [15]). The qualitative behaviour is quite similar, even if the wavelet method gives, in general, lower values. The accuracy in the determination of time variations of the solar cycle lengths is very high for the wavelet method which is completely independent of the arbitrary choice of exact times of sunspots minima or maxima. The long term evolution of the solar cycle length is then confirmed to be in good agreement with the climate changes [15], where high periods are related to weak solar activity and low periods are related to strong solar activity, as well depicted in the most recent cycles.

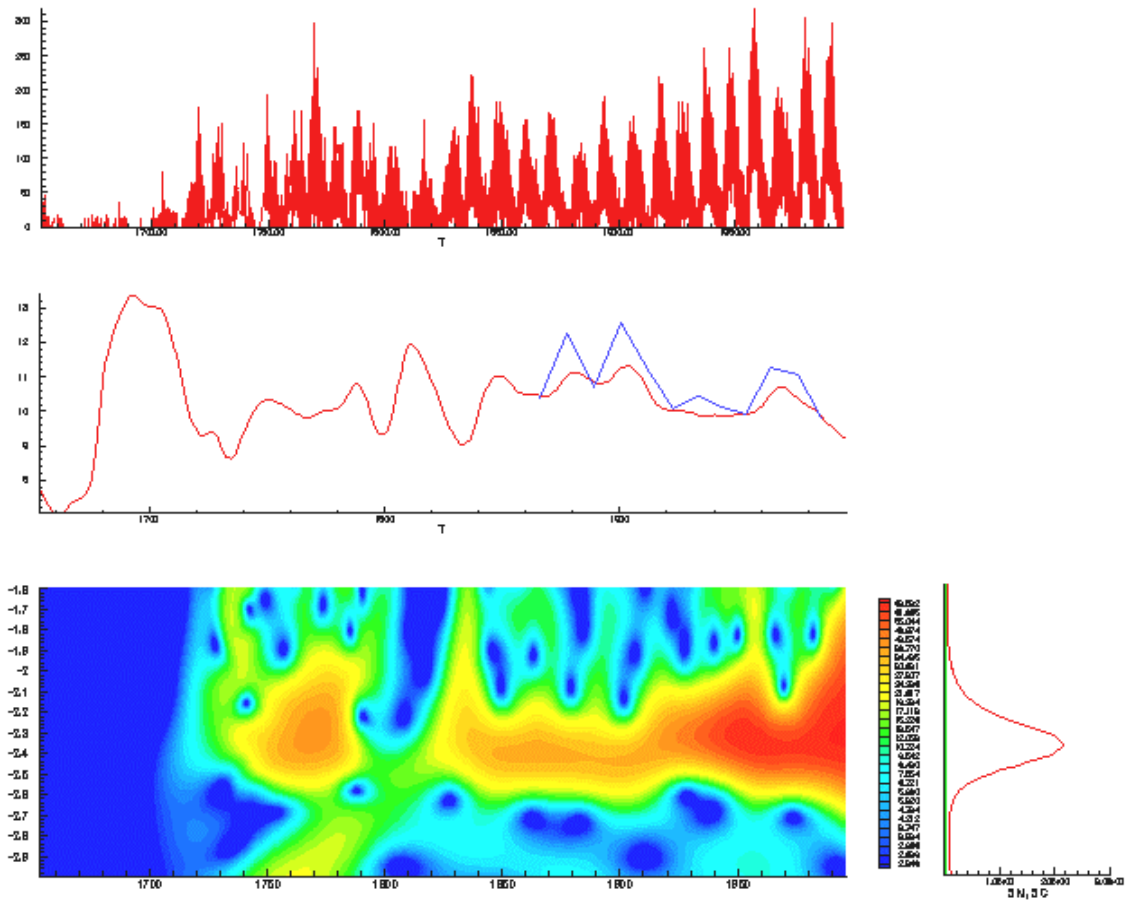

Figure 3

*CONCLUSIONS*

Wavelet analysis approach allows a refined investigation of the temporal variations of solar activity on time scales ranging from days to decades. Considering the daily time series of international sunspot numbers and group sunspot numbers, we analyzed a hierarchy of changing complex periods. In particular, we detected statistically significant periodicities and we excluded a clear evidence of the existence of a near 158 days periodicity correlated to high energy solar flares. A general and accurate determination of the main Schwabe cycle length variations was also suggested on the basis of the wavelet amplitude distribution, extracted from the local wavelet power map.